\documentclass[namedreferences]{SolarPhysics}
\usepackage[optionalrh]{spr-sola-addons} 
\usepackage{graphicx}                    
\usepackage{color}                       
\usepackage{url}                         

\graphicspath{%
    {converted_graphics/}
    {C:/aa/ddr/stereo/HET/}
}
\begin{document}

\begin{article}

\begin{opening}

\title{The Properties of Solar Energetic Particle Event-Associated Coronal Mass Ejections Reported in Different CME Catalogs}
%
\author{I.~G.~\surname{Richardson}$^{1, 2}$\sep
        T.~T.~\surname{von Rosenvinge}$^{1}$\sep
        H.~V.~\surname{Cane}$^3$
       }


%
\runningauthor{I. G. Richardson et al.}
\runningtitle{Properties of CMEs in different catalogs}

%
  \institute{$^{1}$NASA/Goddard Space Flight Center, Greenbelt, Maryland, USA 20771 \\
                     email: \url{ian.g.richardson@nasa.gov}; \url{tycho.t.vonrosenvinge@nasa.gov}\\
             $^{2}$CRESST and Department of Astronomy, University of Maryland, College Park, Maryland, USA 20742\\
             $^3$Bruny Island Radio Spectrometer, Bruny Island, Tasmania, Australia\\
email: \url{hcane@utas.edu.au}\\
$\copyright$ 2015. All rights reserved. 
             }

\begin{abstract}
 
We compare estimates of the speed and width of coronal mass ejections (CMEs) in several catalogs for the CMEs associated with $\sim200$ solar energetic particle (SEP) events in 2006--2013 that included 25~MeV protons.  The catalogs used are: CDAW, CACTUS, SEEDS and CORIMP, all derived from observations by the LASCO coronagraphs on the SOHO spacecraft, the CACTUS catalog derived from the COR2 coronagraphs on the STEREO-A and -B spacecraft, and the DONKI catalog, which uses observations from SOHO and the STEREO spacecraft.  We illustrate how, for this set of events, CME parameters can differ considerably in each catalog.  The well-known correlation between CME speed and proton event intensity is shown to be similar for most catalogs, but this is largely because it is determined by a few large particle events associated with fast CMEs, and small events associated with slow CMEs.  Intermediate particle events ``shuffle" in position when speeds from different catalogs are used.  Quadrature spacecraft CME speeds do not improve the correlation. CME widths also vary widely between catalogs, and are influenced by plane of the sky projection and how the width is inferred from the coronagraph images.  The high degree of association ($\sim50$\%) between the 25~MeV proton events and ``full halo" ($360^o$-width) CMEs as defined in the CDAW catalog is removed when other catalogs are considered.  Using CME parameters from the quadrature spacecraft, the SEP intensity is correlated with CME width, which is also correlated with CME speed.
\end{abstract}

%
\end{opening}

%

\section{Introduction}

Models of particle acceleration at interplanetary shocks predict that the acceleration rate depends, among other parameters, on the speed of the shock through the upstream medium (e.g., \inlinecite{lmg12}, and references therein).  Since many interplanetary shocks that accelerate particles are driven by coronal mass ejections, the widely reported correlation between the intensity of a solar energetic particle (SEP) event and the expansion speed of the related coronal mass ejection (CME) (e.g., \opencite{k78}; \opencite{k84}; \opencite{k87}; \opencite{ream00}; \opencite{gop02}; \opencite{gop04}; \opencite{kv05}; \opencite{c10}; \opencite{r14}; \opencite{lk14}; \opencite{kv14}) is generally interpreted as evidence for particle acceleration by CME-driven shocks.  Nevertheless, the particle intensities associated with CMEs of similar speeds can range over several orders of magnitude even after the influence of the longitude of the solar event relative to the observing spacecraft (e.g.,  \opencite{lar06}; \opencite{l13}; \opencite{r14}) is reduced by selecting events originating in a restricted longitude range.
Factors such as CME width and acceleration, the ambient solar wind speed and ambient SEP intensity (e.g., \opencite{kbr99}; \opencite{k01}), the occurrence of preceding CMEs (e.g., \opencite{gop02}; \opencite{gop04}; \opencite{li12}) and the presence of an interplanetary coronal mass ejection at or beyond the observing spacecraft or between the spacecraft and the Sun \cite{lk14} are among the possible influences on the relationship between SEP intensity and CME speed that have been investigated.  

The CME expansion speed against the plane of the sky is, however, a crude, albeit readily-available, proxy for the shock speed at the point where field lines passing the observer connect with the shock (termed the ``cob-point" by, e.g., \inlinecite{her95} and  \inlinecite{lsh98})  which is most relevant for shock acceleration. Hence, it is not clear that the particle event intensity and CME speed should be closely related. Furthermore, in SEP studies, the CME speed is usually obtained from a catalog of CME parameters.  However, there are several CME catalogs which derive these parameters from coronagraph observations in different ways.  While differences between CME parameters in different catalogs have been discussed previously (e.g.,  \opencite{yas08}; \opencite{rbv09}), the aim of this paper is to reiterate this point in the context of SEP studies.  In particular, we compare parameters in the CDAW, CACTUS, SEEDS and CORIMP catalogs for the CMEs observed by the LASCO coronagraphs on the SOHO spacecraft associated with the $\sim$25 MeV solar proton events identified by \inlinecite{r14}. We also consider CME parameters reported in the DONKI catalog which are inferred from LASCO and STEREO observations in a space weather forecasting environment.  So far as we are aware, this is the first study to examine parameters in the DONKI and CORIMP catalogs for CMEs associated with a large sample of SEP events.  We illustrate examples where the CME parameters deviate considerably in different catalogs and discuss why this is the case.  We also use the CACTUS LASCO and STEREO-A and -B catalogs to obtain the CME speed and width observed by the spacecraft closest in quadrature to the solar event associated with the SEP event, thereby minimizing plane of the sky projection effects.

The various CME catalogs are briefly described in the next section.  In the following sections, CME speeds in the different catalogs are compared, the CME speed--25 MeV proton intensity correlation is examined using speeds from different catalogs,  as are the relationships between CME speed and width, and CME width and SEP intensity.

\section{CME Catalogs}

Many recent studies of the relationship between CME speeds and SEP event intensities (e.g., \opencite{gop02}, \opencite{gop04}; \opencite{kv05};  \opencite{c10}; \opencite{lk14}; \opencite{kv14}; \opencite{r14}) have used parameters from the Catholic University of America/NASA Goddard Space Flight Center ``CDAW" CME catalog (\opencite{y04}; \url{http://cdaw.gsfc.nasa.gov/CME_list/}).  This is compiled manually from observations made by the LASCO C2 and C3 coronagraphs on the SOHO spacecraft \cite{b95}.  The speed inferred from a linear fit to the height-time profile of the {\it fastest feature} on the CME leading edge is most widely used in SEP-related studies.  As with any coronagraph observations, significant plane of the sky projection effects occur in the speed (and also width) if the CME does not originate near the limb of the Sun relative to the observing spacecraft.  Cone models have been developed to help correct for projection (e.g., \opencite{h82}; \opencite{z02}; \opencite{x04}; \opencite{x05}; \opencite{m07}; \opencite{na13};  \opencite{nm14}, and references therein) but have not been widely used in SEP studies (see however, \opencite{pan11}). 

We also consider CME parameters from other LASCO catalogs.  The Royal Observatory of Belgium catalog (\url{http://sidc.oma.be/cactus/}) is compiled using a software package for Computer-Aided CME Tracking (CACTUS) (\opencite{rb04}; \opencite{rbv09}) which estimates the plane of the sky CME speed as a function of position angle, measured counter-clockwise from north.  Both maximum and median speeds based on linear height-time fits to points along the CME front are provided along with the plane-of-the-sky CME width.  CACTUS discards speed measurements that deviate strongly from typical speeds along the CME front.  Thus, the CACTUS and CDAW ``maximum speeds" may differ if the highest speed feature used by CDAW to assess the CME speed is discarded by CACTUS (an example will be shown below).  Furthermore, CACTUS requires a certain number of successive coronagraph images to construct the trajectory of the CME front.  As a result, CMEs faster than $\sim2000$~km~s$^{-1}$ move though the field of view too rapidly for the speed to be estimated, whereas CMEs with speeds $>3000$~km~s$^{-1}$ are occasionally reported in the CDAW catalog.  \inlinecite{yas08} have compared the statistics of the CME properties reported in the CDAW and CACTUS catalogs and, for example, concluded that the properties of many fast CMEs, in particular fast narrow CMEs that are not in the CDAW catalog, are unreliable in the CACTUS catalog.  \inlinecite{rbv09} also compared CDAW and CACTUS LASCO CME parameters and the variation of the CME occurrence rate in each catalog during the solar cycle, also discussed by  \inlinecite{wc14}. 

The SEEDS (Solar Eruptive Event Detection System) catalog (\url{http://spaceweather.gmu.edu/seeds/}) is compiled by applying an automated detection scheme developed at the George Mason University to LASCO C2 coronagraph observations \cite{o08}.  Among other parameters, SEEDS provides linear and second order CME height-time fits.  We also consider the catalog (\url{http://alshamess.ifa.hawaii.edu/CORIMP/}) developed by the University of Hawaii, Aberystwyth University and Trinity College, Dublin using the CORIMP (coronal image processing) method (\opencite{b12}; \opencite{m12}). 

CMEs are also observed by the COR2 coronagraphs \cite{h08} on the twin STEREO-A (``Ahead") and -B (``Behind") spacecraft, launched on October 26, 2006 and placed into heliocentric orbits advancing ahead of or lagging the Earth in its orbit, respectively.  We use STEREO CME identifications made using CACTUS, available at \url{http://sidc.oma.be/cactus/}.  Note that the thresholds used by CACTUS when detecting the CME signal using the Hough transform are different for LASCO and STEREO-A (0.3) and STEREO-B (0.25), so the identification criteria in these catalogs are not completely equivalent. 

We will also use CME parameters from the ``Space Weather Database Of Notification, Knowledge, Information (DONKI)" (\url{http://swc.gsfc.nasa.gov/main/donki}) developed at the Community Coordinated Modeling Center (CCMC), NASA Goddard Space Flight Center.  DONKI includes estimates of CME speed, width and propagation direction.  Since these are obtained in a forecasting environment using available real-time observations, the methods used may vary from event to event depending on which observations are available.  CME speeds may be inferred from LASCO and STEREO coronagraph observations, if available, by triangulation \cite{liu10} using the STEREO-CAT tool developed by CCMC (\opencite{l14}; \url{http://ccmc.gsfc.nasa.gov/analysis/stereo/}) or the SWPC-CAT tool developed by the Space Weather Prediction Center \cite{m13}.   In other cases, single spacecraft observations are used.  These estimates, available for a selection of CMEs observed since June 2010, are used by CCMC as input to the WSA-ENLIL+Cone CME propagation model to provide a forecast of CME arrival for spacecraft operators (\opencite{z13}; \opencite{l14}). Since the WAS-ENLIL+Cone CME parameters are input at 21.5~$R_s$, they are estimated at this height, taking into account, for example, the observed CME speed and acceleration within this distance from the Sun.  

\begin{figure}
\centerline{\includegraphics[width=4.0in,height=6.0in,keepaspectratio]{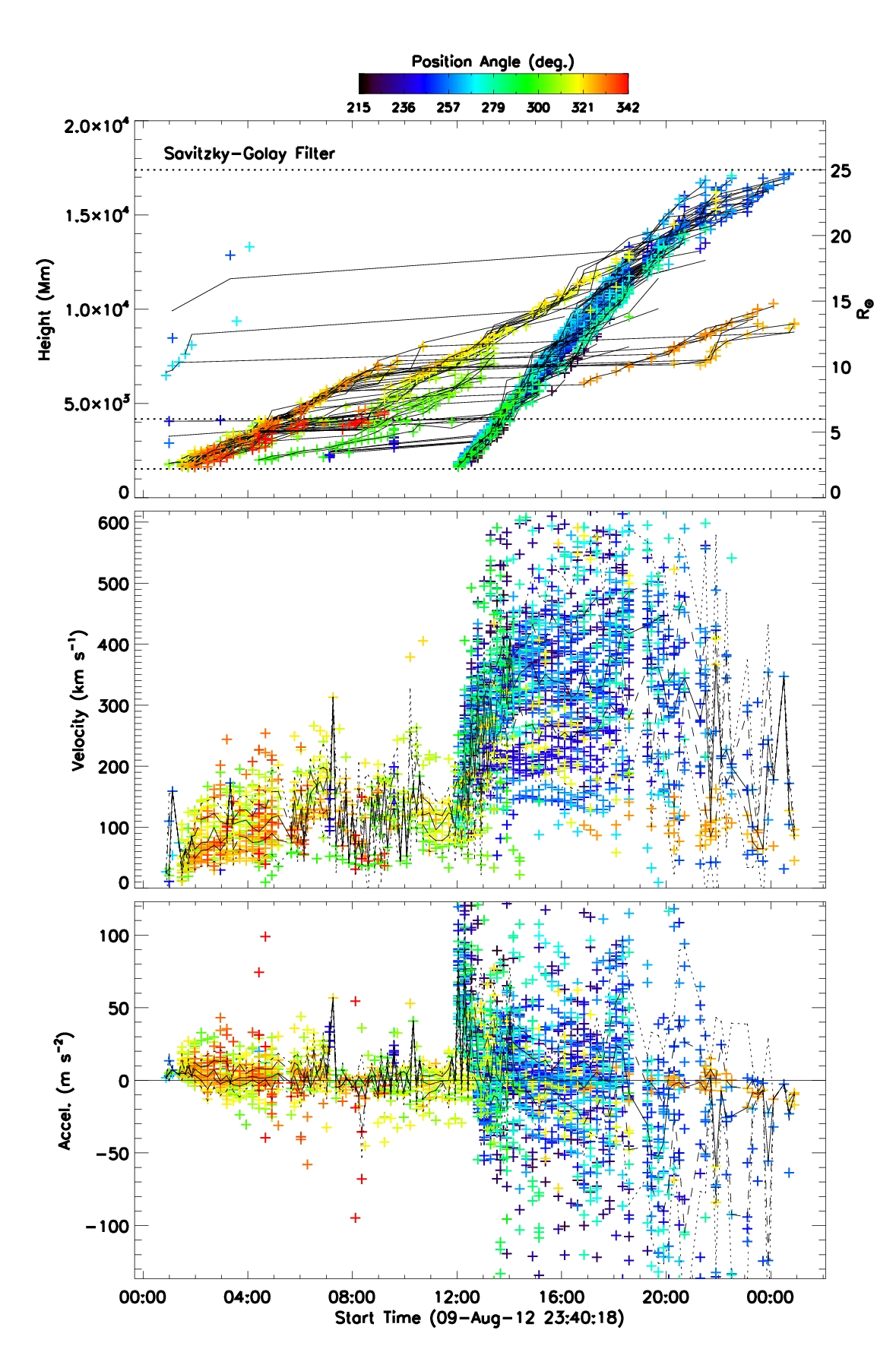}}
\caption{Observations for a CORIMP CME catalog entry (first detection time August 10, 2012; 0052:24 UT) that encompasses more than one CME.  The panels show the height-time profiles of features identified by a Savitsky-Golay filter at position angles indicated by the color code, the speeds of these features, and their acceleration.  A slow CME originating around 02 UT and a faster CME from around 12 UT that is associated with an SEP event observed at STEREO-A and the Earth, are evident.}
\label{corimp}         
\end{figure}

We will compare the parameters in these catalogs for the CMEs associated with around 200 solar energetic particle events including 25~MeV protons observed at STEREO and/or at Earth since STEREO launch  \cite{r14}.  In most cases, the ``first detection time" of a CME is consistent with the onset time of the related SEP event.  However, a single CME entry in the CORIMP catalog may include several individual CMEs evident in LASCO movies and the CME height--time and speed profiles provided in the catalog.  Figure~\ref{corimp} (from the CORIMP on-line catalog) shows as an example, observations related to the CORIMP CME catalog entry with a first detection time of August 10, 2012; 0052:24 UT.  This is the only entry on this day in the catalog, and the CME analysis period extends to the end of the day. The top panel shows the height-time profiles of features identified by a Savitsky-Golay filter at position angles indicated by the color scale.  The center panel gives the speeds of these features, and the bottom panel, their acceleration.  This interval includes a slow CME originating around 02 UT and a faster CME from around 12 UT that is associated with a 25 MeV proton event observed at STEREO-A and Earth \cite{r14}.  The catalog gives  ``median" and ``maximum" CME speeds of 175 km~s$^{-1}$ and 375 km~s$^{-1}$ but these clearly do not characterize the CME associated with the SEP event, which includes features with speeds up to around 600 km~s$^{-1}$ (Figure~\ref{corimp}).  In such cases where multiple CMEs are present in a CORIMP interval, we have estimated the speed of the specific CME of interest from the highest speed features evident in the speed--time plots rather than use the speed quoted in the CORIMP catalog.

\begin{table}
\renewcommand{\arraystretch}{.7}
\setlength{\tabcolsep}{.02in}
\caption[]{CME Parameters for January 23, 2012 CME}
\label{comp} 
\begin{tabular*}{\maxfloatwidth}{l|cccc|c|c|c|} 
\hline
\multicolumn{1}{c|}{ }&\multicolumn{4}{c|}{SOHO-LASCO}&\multicolumn{1}{c|}{STA COR2}&\multicolumn{1}{c|}{STB COR2}&\multicolumn{1}{c|}{ }\\
                       &CDAW      &CACTUS &SEEDS    & CORIMP     &CACTUS     &CACTUS &DONKI\\
\hline
Onset (U.T.)          &0400    &0436  & 0348      &$0000^a$        &$0524^b$                 &0354        &0400     \\
(Max.) Speed (km~s$^{-1}$)   & 2175      & 2016 &  700         &$1376^a$      &   $1136^b$              &   1785    &  2211     \\
Median Speed (km~s$^{-1}$) &...         &1092 &  ...          &$557^a$       &   $556^b$              & 1136      &  ...     \\
Width ($^o$)          &360        & 360  & 183          &$116^a$       &   $210^b$              & 352        & $124^c$ \\

\hline 
\end{tabular*}
$^a$CME interval includes preceding structures; $^b$The STEREO-A CACTUS catalog identifies a CME that has a position angle $\sim180^o$ away from that of the actual CME; $^c$Twice the half width given in the DONKI catalog.
\end{table}

\begin{figure}    
\centerline{\includegraphics[width=1.0\textwidth,clip=]{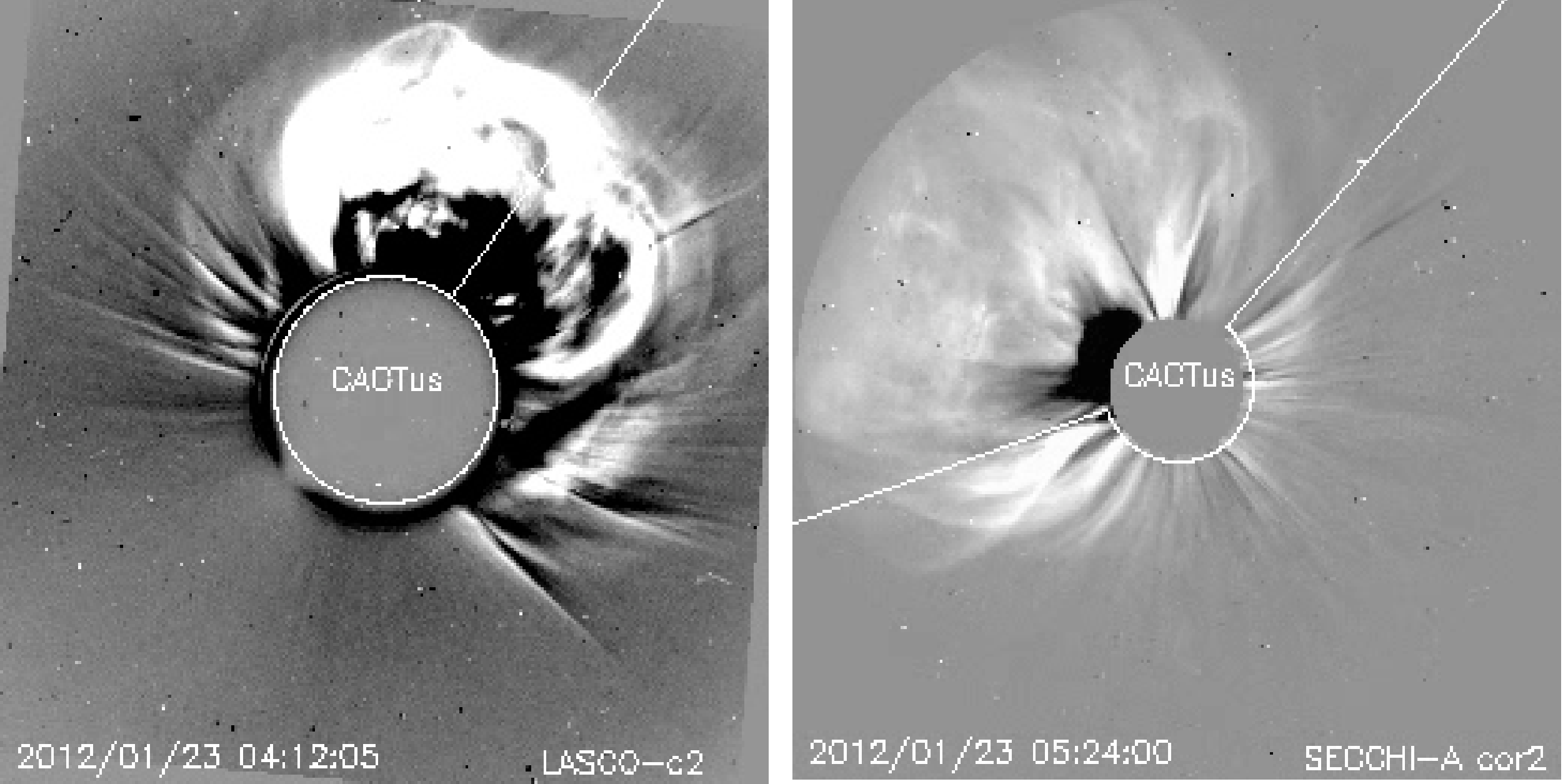}}
              \caption{The CME at $\sim04$~UT on January 23, 2012 from a solar event at W$21^o$ as viewed by LASCO near the Earth (left) and STEREO-A  COR2 at $87^o$ west of the solar event (right); images are from the CACTUS catalog.  Both the CDAW and CACTUS catalogs identify the LASCO CME as a ``full halo" ($360^o$-width).  At STEREO-A, CACTUS identifies a ``fictitious" wide CME moving to the southwest (the white arc at the edge of the occulting disk indicates the extent of the CME).}
   \label{jan23CME}
   \end{figure}

Table~\ref{comp} illustrates the range of CME parameters found in the different catalogs for another example CME, on January 23, 2012,  associated with an M8.7 flare at W21$^o$ relative to Earth and a large 25 MeV proton event commencing at $\sim04$~UT \cite{r14}. Example LASCO and STEREO-A COR2 images (from the CACTUS catalog) are shown in Figure~\ref{jan23CME}.  Four of the seven catalogs agree that this was a fast CME with a speed of $\sim2000$~km~s$^{-1}$, whereas SEEDS gives a speed of only 700~km~s$^{-1}$.  Similar to the case discussed above, the CORIMP LASCO analysis interval (which starts around four hours ahead of the actual CME onset) includes slower preceding structures that are not associated with the CME of interest.  The CORIMP speed-time plot for this event indicates structures within this CME with speeds approaching $\sim2000$~km~s$^{-1}$, though the CME speeds quoted in the catalog for the more extended interval (Table~\ref{comp}) are lower.  Interestingly, the STEREO-A CACTUS catalog identifies a ``fictitious" CME (right-hand panel of Figure~\ref{jan23CME}) moving in the opposite direction to the actual CME (the white arc at the edge of the occulting disk indicates the extent of the CME identified by CACTUS).  The parameters of this ``CME" are given in Table~\ref{comp}.  Both the CDAW and LASCO CACTUS catalogs classify this as a full halo (360$^o$ width) CME even though it was clearly asymmetric and predominantly moving to the north west. STEREO-B CACTUS suggests a similar wide ($352^o$) CME, while other catalogs give widths of 116--183$^o$.

\section{Results}
\subsection{CME speeds}

\begin{figure}    
 \centerline{\includegraphics[width=1.0\textwidth,clip=]{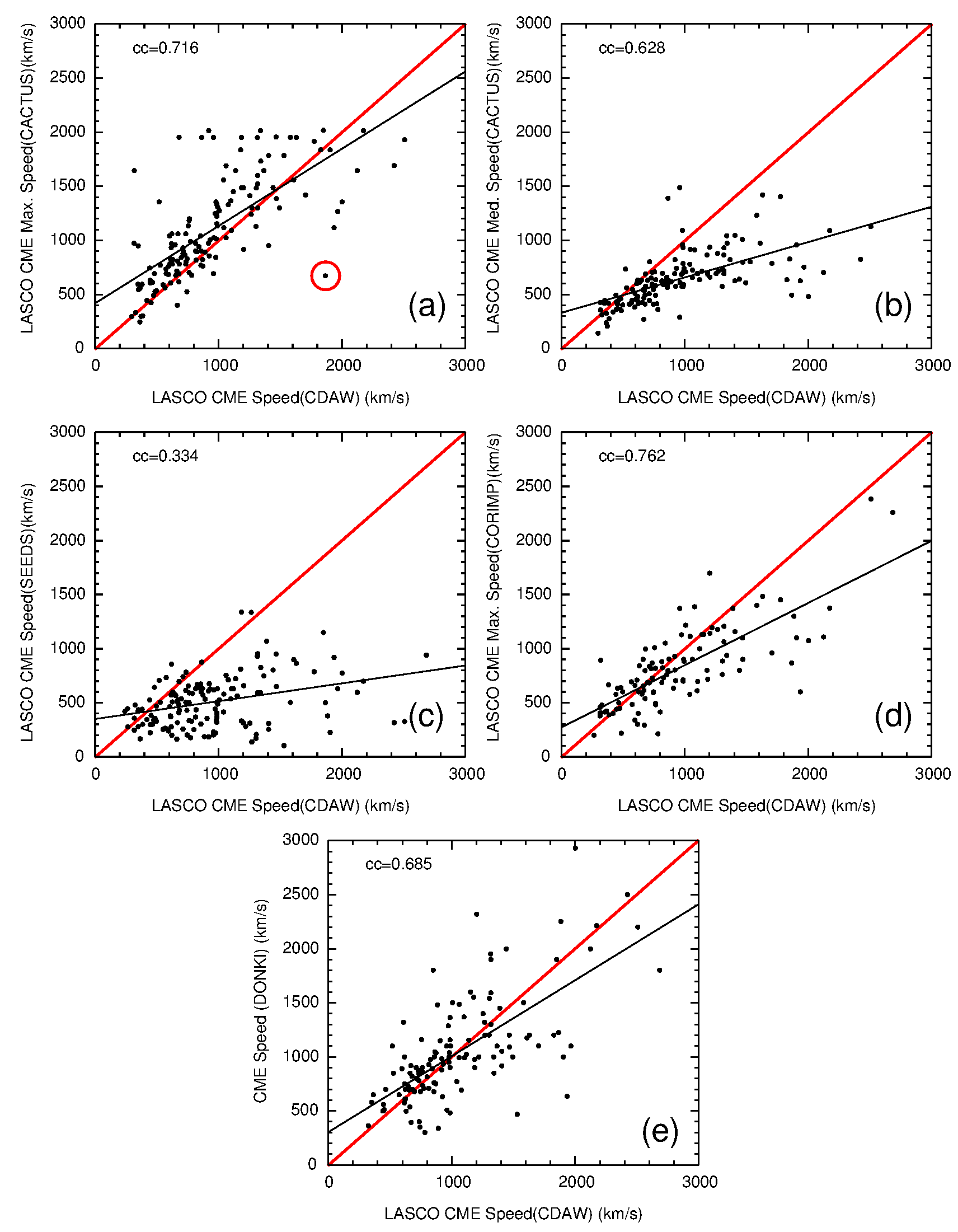}}
              \caption{Comparison of linear-fit CME speeds from the CDAW LASCO catalog, for CMEs associated with 25 MeV proton events in December 2006 to May 2013, with (a) the LASCO CACTUS maximum speed, (b) the LASCO CACTUS median speed, (c) the LASCO SEEDS speed, (d) the LASCO CORIMP speed, and (e) the speed from the DONKI catalog.  In each panel, the black line indicates the linear least squares fit, while the red line indicates equal values.  The red circle in (a) indicates the event illustrated in Figure 4.}
   \label{cdctv}
   \end{figure}

\begin{figure}    

\includegraphics[width=2.3in,height=2.3in,keepaspectratio]{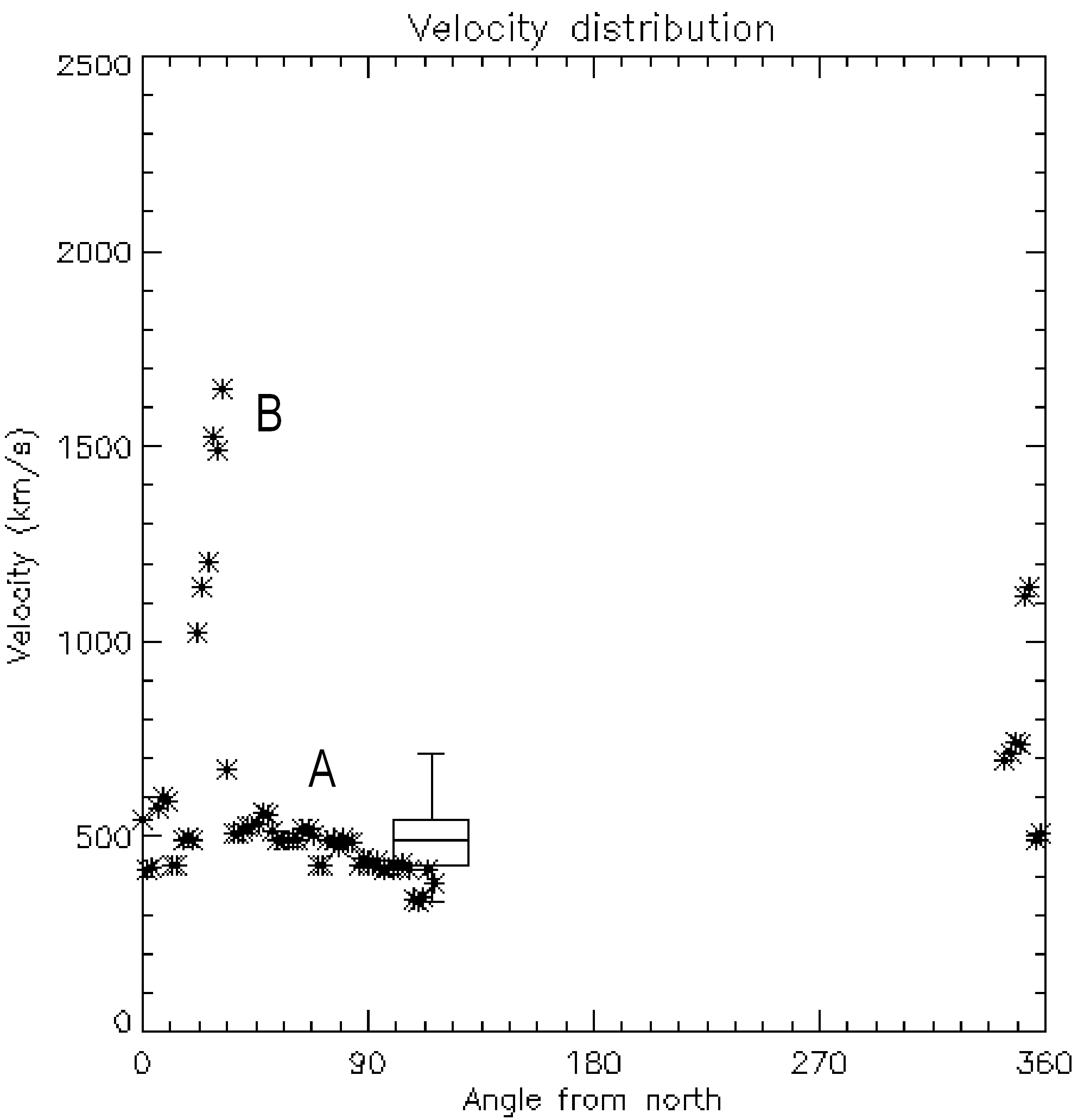}
\includegraphics[width=2.3in,height=2.3in,keepaspectratio]{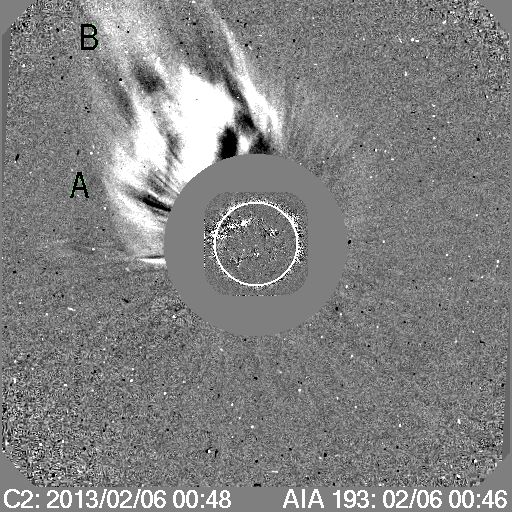}        
              \caption{CACTUS LASCO CME speed plotted against position angle from north (left) for the 0024~UT CME on February 6, 2013 moving to the north east (right).  The CDAW catalog speed (1867 km~s$^{-1}$) is based on the narrow fast-moving feature (B) moving ahead of the main CME front (A), whereas CACTUS gives a maximum speed of 672 km~s$^{-1}$, more representative of the broad CME front, because points exceeding three times the standard deviation of the speeds along the CME indicated by the error bars in (a), including the narrow high-speed flow, are excluded.}

  \label{cac130206}
   \end{figure}

We first compare the CME speeds in various catalogs with the linear-fit speeds from the CDAW LASCO catalog. Figure~\ref{cdctv}(a) shows ``maximum speeds" from the CACTUS LASCO CME catalog for a sample of 145 CMEs associated with 25 MeV proton events in December 2006 to May 2013 (the most recent CDAW parameters available at the time of the analysis).  The speeds are well correlated ($cc=0.716$), but there is considerable scatter, and the CACTUS maximum speeds generally exceed CDAW speeds (the red line indicates equality) for this set of events.  The cut-off in CACTUS speeds at $\sim2000$~km~s$^{-1}$ is evident together with a cluster of events with speeds near this limit.  For several events, the CACTUS and CDAW LASCO speeds are quite different.  The example circled, at 0024~UT on February 6, 2013, associated with a C8.7 flare at N22$^o$E19$^o$ relative to Earth, has a speed of 1867 km~s$^{-1}$ in the CDAW catalog, but only 672 km~s$^{-1}$ in the CACTUS catalog.  Figure~\ref{cac130206} (from the CACTUS catalog) shows the speed inferred by CACTUS as a function of position angle from north for this CME together with a representative LASCO image.  The CDAW speed appears to be based on the narrow high-speed flow (B) moving ahead of the main CME front (A) whereas CACTUS only accepts points within three standard deviations of the mean speed along the CME front, indicated by the error bars in the left-hand panel, and discards the high speed flow, reporting a speed that is more representative of the main CME front.  Similar circumstances occur for other outliers in Figure~\ref{cdctv}(a). Including events with speeds below 1800~km~s$^{-1}$ in both catalogs, removing the outliers and events close to the CACTUS cut off, increases the correlation coefficient to 0.77.

Figure~\ref{cdctv}(b) shows that CACTUS LASCO ``median" speeds are also correlated with, but as expected, are generally lower than, the CDAW speeds ($cc=0.727$).  We include this figure to contrast with the dependence of the maximum speeds in (a).   We note that \inlinecite{r14} used the median CACTUS LASCO CME speed in their event table for those CMEs for which no CDAW speed was available, but Figures~\ref{cdctv} (a) and (b) indicate that the maximum CACTUS speed would have been a more comparable choice.

Figure~\ref{cdctv}(c) shows the poorer correlation between CDAW and SEEDS LASCO CME speeds ($cc=0.334$).  With a few exceptions, the SEEDS speeds are under 1000 km~s$^{-1}$ and lie below the CDAW speeds. \inlinecite{o08} show a similar figure (their Figure~7(c)) for a different CME sample.  Figure~\ref{cdctv}(d) shows CORIMP LASCO maximum speeds.  There are two CMEs with speeds above 2000 km~s$^{-1}$ in the CDAW and CORIMP catalogs.  In both cases, CORIMP identified the individual CMEs associated with the SEP events, and these speeds are taken from the catalog.  Speeds are comparable at low speeds, though with much scatter, while above $\sim1000$~km~s$^{-1}$ CORIMP LASCO speeds generally fall below CDAW speeds.  Figure~\ref{cdctv}(e) shows the correlation between DONKI and CDAW speeds ($cc=0.685$).  DONKI speeds are spread over several hundred km~s$^{-1}$ for a given CDAW speed and lie on both sides of the line of equality, indicating no clear bias with respect to the CDAW speeds.

\begin{figure}    
\centerline{\includegraphics[width=1.0\textwidth,clip=]{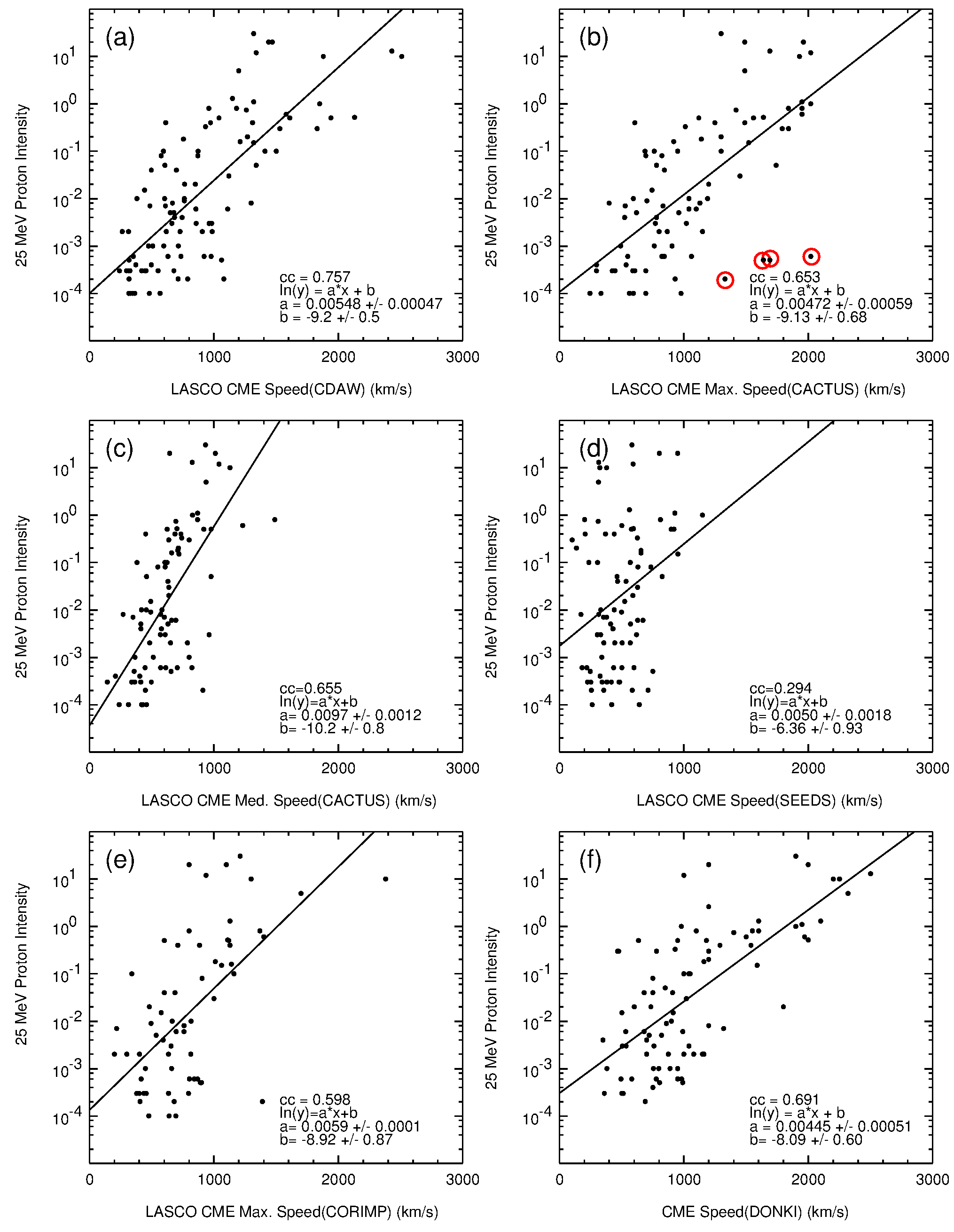}}
              \caption{Intensity of 25~MeV protons (Richardson {\it et al.}, 2014) plotted \textit{vs.} the speed of the associated CME for events 30--$90^o$ west of the particle-observing spacecraft. (a) CDAW LASCO CME speed; (b) CACTUS LASCO maximum CME speed; (c) CACTUS LASCO median CME speed; (d) SEEDS LASCO CME speed; (e) CORIMP LASCO CME speed; and (f) DONKI CME speed.}
   \label{sepspeed}
   \end{figure}

\subsection{Correlation of CME speeds with SEP intensity}

We now illustrate how using CME speeds from different catalogs influences the widely-discussed correlation between CME speed and SEP intensity, specifically the peak 25~MeV proton intensity for the events identified by \inlinecite{r14}. As discussed in that paper, this intensity is taken early in the event rather than at the passage of the related interplanetary shock, if present.

Figure~\ref{sepspeed} shows the 25~MeV proton intensity versus CME speed from the different CME catalogs for ``well-connected" events at W$30^o$ to W$90^o$ relative to the observing spacecraft, which may be STEREO-A/B or SOHO.  For example, a CME with an eastern hemisphere source relative to Earth may however be within this western longitude range relative to STEREO-B located above the east limb of the Sun; the particle intensity is then measured at STEREO-B.  Figure~\ref{sepspeed}(a) shows the distribution obtained using CDAW speeds.  The usual positive correlation is evident, with a correlation coefficient $cc=0.757$.  (Though we quote correlation coefficients to help compare the distributions in Figure~\ref{sepspeed}, there is however no physical reason to expect the log intensity and CME speed to be linearly correlated.) 

\begin{figure}    
\centerline{\includegraphics[width=1.0\textwidth,clip=]{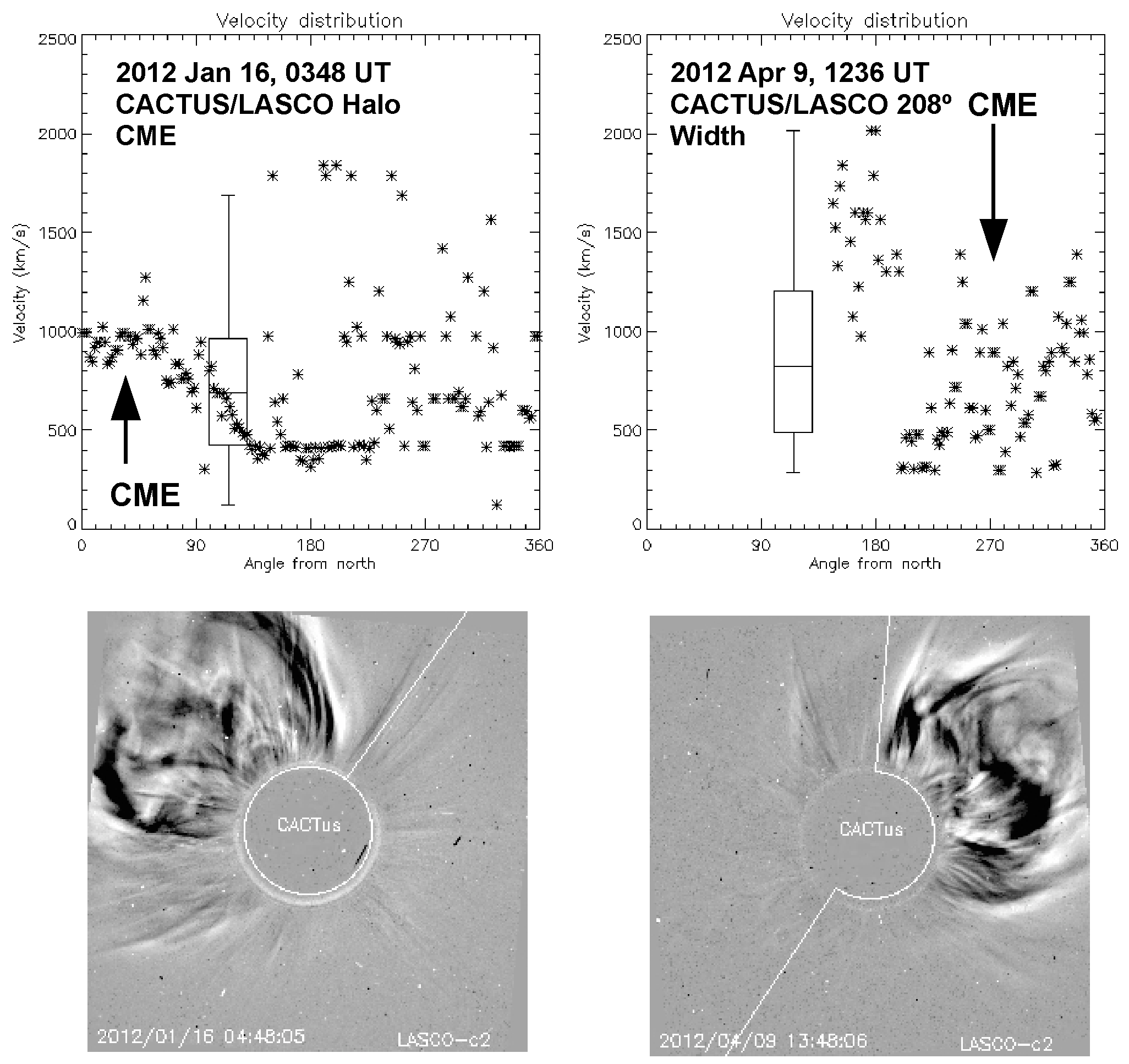}}
              \caption{LASCO CME speed as a function of position angle from CACTUS for CMEs on (left) January 16, 2012, and (right) April 9, 2012, two of the events circled in red in Figure~5(b) for which an apparently fast CME in the CACTUS LASCO catalog is associated with a weak 25~MeV proton event.  In both cases, CACTUS identifies a CME that is more extended (indicated by the arc surrounding the occulter) than is suggested by the coronagraph observations and selects maximum CME speeds based on spurious high-speed flows outside the visible CME front.  In contrast, the CDAW catalog speeds (1060 km~s$^{-1}$, January~16, left) and 921 km~s$^{-1}$ (April 9, right) are more representative of the speeds of the CME fronts, and more consistent with the weak particle events.  }
   \label{20120104}
   \end{figure}

A positive correlation ($cc=0.653$) is also found using CACTUS LASCO maximum speeds in (b).  There are, however, four fast CMEs (circled) apparently associated with unexpectedly weak SEP events.  CACTUS speed versus position angle plots and sample LASCO C2 coronagraph images are shown for two of these cases in Figure~\ref{20120104}.  The January 16, 2012 CME (left) moved to the northeast.  However, CACTUS identifies this as a ``full halo" ($360^o$ width) CME and determines the CME speed from observations at all position angles, including apparently spurious fast flows in the west, far outside the CME, that result in CACTUS assigning a high maximum speed of 1689 km~s$^{-1}$.  In contrast, the CDAW catalog gives a speed of 1060 km~s$^{-1}$ that is consistent with CACTUS speeds along the CME front in the northeast.  Similarly, for the April 9, 2012 event (right), the CME moved to the northwest but CACTUS infers a width of $208^o$ that extends well beyond the prominent CME front.  CACTUS again identifies a high maximum speed (2016 km~s$^{-1}$) based on spurious fast flows in the south east, whereas the CDAW catalog gives a speed of 921 km~s$^{-1}$ that is reasonably representative of the (variable) speed of the CME front.  Adopting the CDAW speeds, these events are no longer outliers to the distribution (\textit{cf.}, Figure~\ref{sepspeed}(a)).  Note that in these cases, in contrast to the event in Figure~\ref{cac130206}, it is the CDAW catalog rather than CACTUS that provides the more reasonable estimates of the LASCO CME speeds.  

Figure~\ref{sepspeed}(c) shows a correlation ($cc=0.655$) between CACTUS LASCO median CME speeds and 25~MeV proton intensity.  The best fit line is steeper due to the narrower CME speed range.  Figure~\ref{sepspeed}(d) shows the weak correlation of SEEDS speed with the 25~MeV proton intensity ($cc=0.294$), whereas stronger positive correlations are found using CORIMP (e; $cc=0.598$) and DONKI (f; $cc=0.691$) CME speeds.  

Note that, comparing the fits in Figure~\ref{sepspeed}(a), (b), (e) and (f), the proton intensity--CME speed correlation is similar whether CDAW, CACTUS or CORIMP LASCO CME speeds, or speeds from DONKI are used.  The reason for this is that the fits are largely determined by the proton intensity which spans $\sim$six orders of magnitude whereas the CME speeds extend $\lsim$ one order of magnitude, so that the overall fits do not depend strongly on which catalog provides the CME speed.  Choosing different catalogs for the CME speeds causes the points to ``shuffle around" as different speeds are inferred for individual events without influencing the overall positive correlation with proton intensity.

\begin{figure}    
  \centerline{\includegraphics[width=1.0\textwidth,clip=]{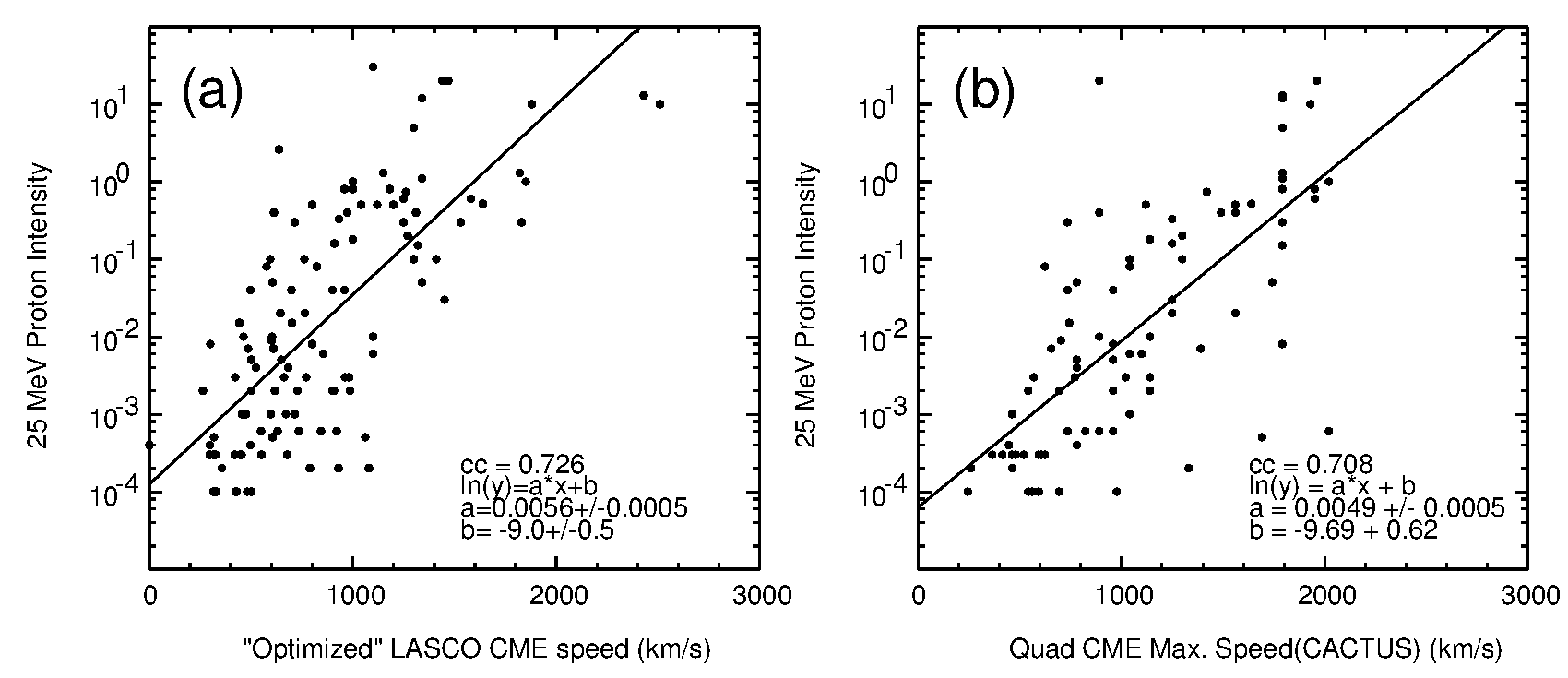}}
              \caption{25 MeV proton intensity plotted against (a) an ``optimized" LASCO CME speed selected from the CDAW or CACTUS catalogs, and (b) the CACTUS CME speed observed by the spacecraft in near-quadrature to the related solar event.}
   \label{opt}
   \end{figure}

We have considered two ways in which the CME speed might be assessed more reliably and examined the effect on the proton intensity--speed distribution.  Above, we discussed cases (Figures~\ref{cac130206} and \ref{20120104}) where either the CDAW or CACTUS LASCO maximum speed was more representative of the CME speed based on examination of the CACTUS LASCO speed versus position-angle plots.  Thus, for each LASCO CME in this study, we used the speed--position angle plots together with coronagraph images/movies (to verify the range of position angles occupied by the prominent CME structures) to infer whether the CDAW or CACTUS LASCO maximum speed better represents the speed of the CME.  Figure~\ref{opt}(a) shows the proton intensity plotted against this ``optimized" LASCO CME speed.  The correlation coefficient ($cc=0.726$) and fit are both comparable to most of those in Figure~\ref{sepspeed}. 

We have also used CACTUS CME speeds obtained from the spacecraft (SOHO or STEREO-A or B) located closest to quadrature, $\sim90^o$ longitude from the solar event that gave rise to the CME and SEP event (solar event longitudes are listed in \inlinecite{r14}).  At this location, plane of the sky projection effects on CME speeds should be minimized.  In practise, we considered observations made by spacecraft $60^o$--$120^o$ east or west from the longitude of the solar event, choosing the spacecraft closest to $90^o$ separation if more than one was within this range.  Although we do include observations from different spacecraft, as mentioned above, the STEREO-A/B and LASCO CACTUS CME identifications use different thresholds so the CME speeds inferred may not be completely equivalent.  Figure~\ref{opt}(b) shows that the correlation and fit between the CACTUS maximum CME speed from the quadrature spacecraft and 25~MeV proton intensity are similar to those obtained in Figure~\ref{sepspeed} and \ref{opt}(a).  Some of the outliers in Figure~\ref{sepspeed}(b) discussed above are also evident in this distribution.  Thus, Figure~\ref{opt} indicates that choosing what might be expected to be a more reliable estimate of the CME speed does not substantially change the correlation between SEP intensity and CME speed, though the speeds for individual events do change.

\begin{figure}    
  \centerline{\includegraphics[width=1.0\textwidth,clip=]{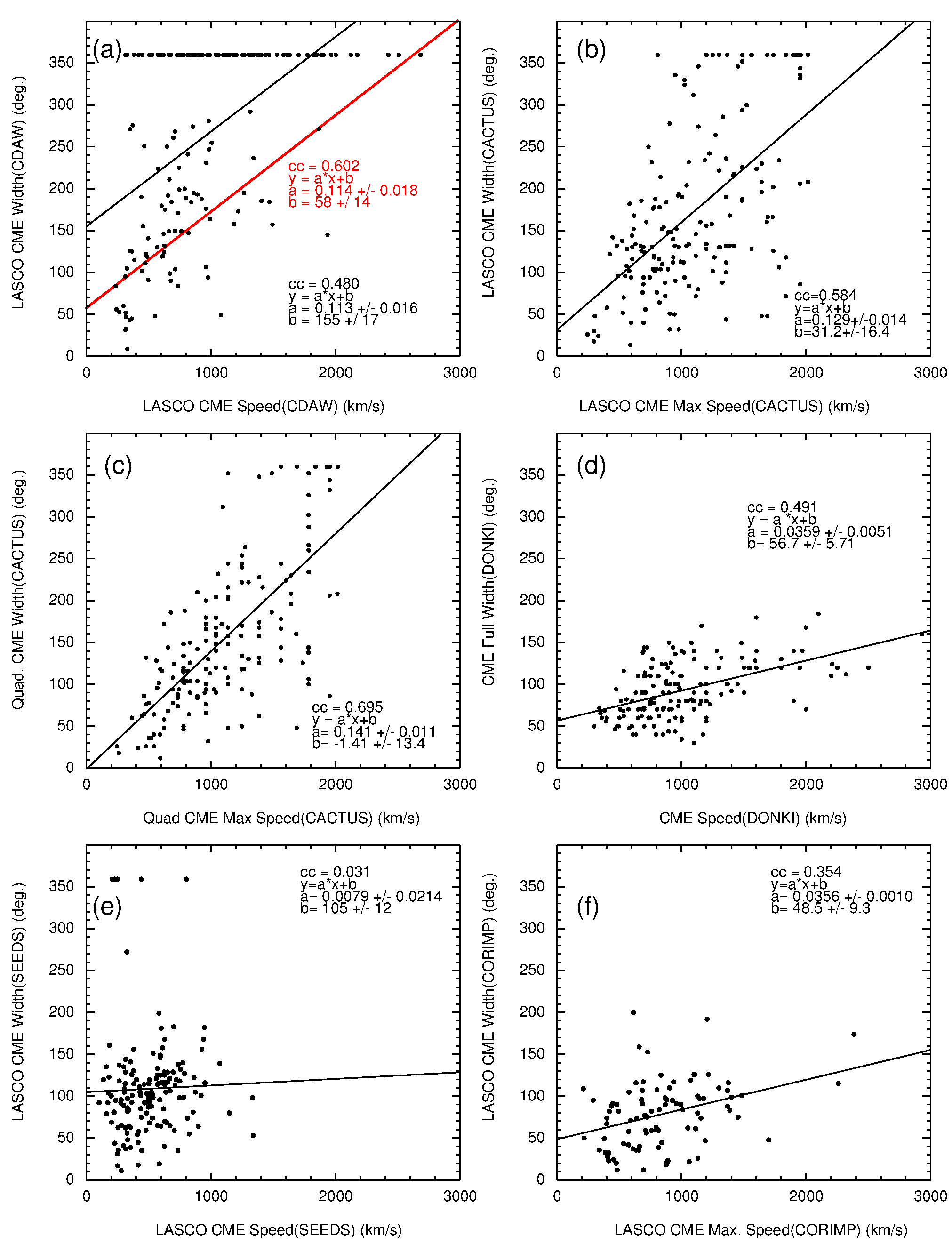}}
              \caption{Comparison of CME widths versus speeds (for CMEs associated with the 25 MeV proton events of Richardson et al. (2014)) for LASCO in the (a) CDAW and (b) CACTUS catalogs, (c) for CACTUS observations at the quadrature spacecraft, (d) from the DONKI catalog (doubling the half widths given in this catalog), (e) from LASCO/SEEDS and (f) from LASCO/CORIMP.  In (a), 51\% of the events are ``full halo" CMEs; the red line indicates the fit to events with speeds $<1600$~km~s$^{-1}$ and widths $<250^o$.  }
   \label{widthspeed}
   \end{figure}

\subsection{CME widths}

As noted in the introduction, some studies have considered whether the SEP intensity is also related to CME width.  However, CME widths are also strongly influenced by projection, and how the different catalogs assess the width.   Furthermore, adjacent pre-existing structures, shocks, and compressions may be included in the width in addition to the actual CME.  Comparing CME widths in the CDAW and CACTUS LASCO catalogs,  \inlinecite{rbv09} concluded that beyond $\sim120^o$, the width is not well-defined, in particular for halo CMEs.  Observations from multiple spacecraft have been used to infer the CME shape (e.g., \opencite{ther11}, \opencite{theret11}, \opencite{liew11}, \opencite{bos12}, \opencite{r12},  \opencite{col13}) but only for a limited number of cases.  Here we will consider the widths of the CMEs associated with the $\sim25$~MeV proton events of \inlinecite{r14} as reported in the various CME catalogs.

We first illustrate in Figure~\ref{widthspeed} the highly-variable relationship between CME width and speed reported in these catalogs.  In the CDAW catalog (a), some 51\% of these CMEs are classified as ``full halo" ($360^o$) CMEs.  The close association between CDAW halo CMEs and SEP events has been noted previously (e.g., \opencite{cro98}; \opencite{rco99}; \opencite{gop02};  \opencite{c10}).  These CMEs are however, frequently highly asymmetric and may originate far from central meridian -- the CDAW full halo CMEs considered here originated at all longitude ranges relative to the Earth, including near the limbs. The red line shows the correlation ($cc=0.602$) when CMEs with speeds less than 1600~km~s$^{-1}$ and widths less than $250^o$ are considered, suggesting that faster CMEs tend to be wider.  (Again, we give the correlation coefficients and fits not because a linear correlation between CME speed and width is expected, but to compare the distributions in the different catalogs.)  A fit for all points ($cc=0.480$) is also shown.  

In the CACTUS LASCO catalog (Figure~\ref{widthspeed}(b)) the fewer (13\%) full halo CMEs are predominantly associated with faster (maximum speed $>1200$~km~s$^{-1}$) CMEs.  Figure~\ref{widthspeed}(c) shows CACTUS CME parameters from the spacecraft in quadrature to the solar event, suggesting a correlation between maximum CME speed and width ($cc=0.695$).  Only 4\% of these CMEs are classified as full halos.  Turning to the DONKI database (Figure~\ref{widthspeed}(d)), we first double the CME half-widths provided, to correspond to the full widths provided in the other catalogs.  DONKI shows a weaker correlation between width and speed ($cc=0.491$), apparently resulting from an absence of fast but narrow CMEs.  DONKI CME widths rarely exceed $150^o$ even though they are generally inferred from plane of the sky observations.  However, if multiple spacecraft observations are available, the narrowest width observed is used in the catalog.

SEEDS CME speeds and widths (Figure~\ref{widthspeed}(e)) are uncorrelated ($cc=0.031$), while CORIMP (f) shows a weaker correlation ($cc=0.354$) than some other catalogs. CORIMP also does not identify particularly wide CMEs.  Widths rarely exceed $\sim150^o$ and no full halo CMEs are identified even when a clear symmetrical halo CME is observed, such as the event on February 15, 2011 (onset time 0224~UT from the CDAW catalog) identified as a full halo by CDAW and CACTUS. A width of only $200^o$ is assigned by CORIMP (the event also has an earlier start time of February 14, 2212~UT in this catalog).  Similarly, SEEDS only identifies a $67^o$-width sector of this halo CME.

\begin{figure}    
 \centerline{\includegraphics[width=1.0\textwidth,clip=]{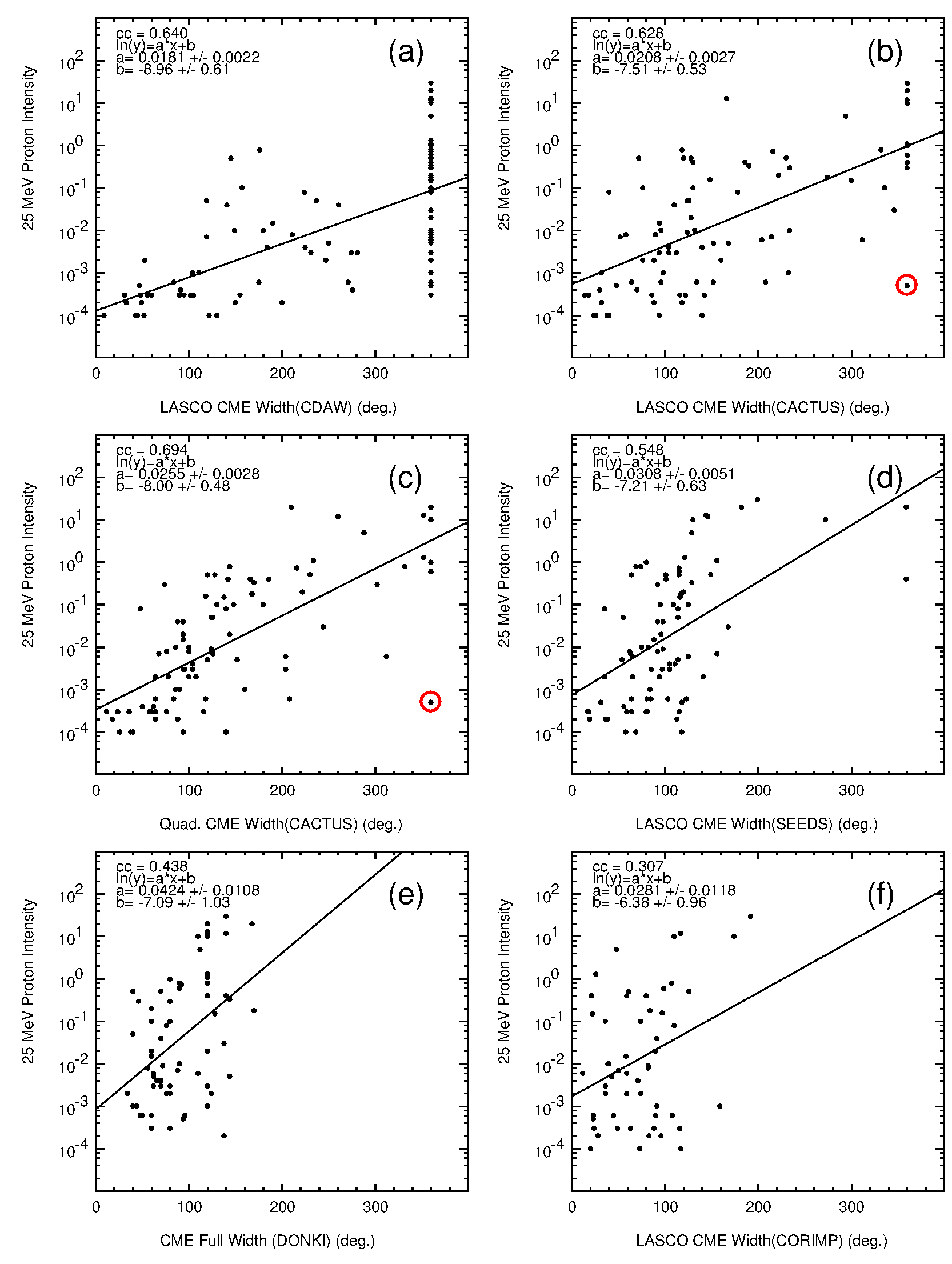}}
              \caption{Correlations of the 25~MeV proton intensity with the CME width in various catalogs, for events 30--90$^o$ west of the observing spacecraft.  The event circled red is apparently a weak proton event associated with a full halo CME in the CACTUS LASCO catalog.  However, the width of this asymmetric CME may be narrower than estimated.}
   \label{CMEwidthsep}
   \end{figure}

We now consider the relationship between the $\sim25$~MeV proton intensity and CME width in the various catalogs, as illustrated in Figure~\ref{CMEwidthsep} for events originating between 30 and 90$^o$ west of the observing spacecraft. Figure~\ref{CMEwidthsep}(a) uses CME widths from the CDAW catalog and again is dominated by events associated with full halo CMEs which span nearly the full range of SEP intensities.  There is a near-absence of 25 MeV proton events exceeding $\sim10^{-3}$~(MeV cm$^2$ s sr)$^{-1}$ for CMEs narrower than $\sim100^o$.  A similar tendency for narrow ($\sim80^o$-width) CMEs to be associated with weak SEP events (below $\sim10^{-3}$~(MeV cm$^2$ s sr)$^{-1}$) is also evident using CACTUS LASCO CME widths (Figure~\ref{CMEwidthsep}(b)) though there are exceptions.  Overall, there is a positive correlation between CME width and SEP intensity ($cc=0.628$), with the few full halo CMEs being associated with large SEP events.  The apparent exception (circled), on January 16, 2012, also identified as a CDAW full halo CME, is however asymmetric and largely confined to the north-east quadrant in LASCO images, so the width may be overestimated in the CACTUS and CDAW LASCO catalogs.

Figure~\ref{CMEwidthsep}(c) shows CACTUS CME widths from the spacecraft closest in quadrature to the solar event.  Overall, there is a clear correlation between CME width and 25~MeV proton intensity ($cc=0.694$).  Again narrow CMEs tend to be associated with weak SEP events -- CMEs with widths out to $\sim60^o$ are associated with SEP events with 25~MeV proton intensities that rarely exceed $\sim10^{-3}$~(MeV cm$^2$ s sr)$^{-1}$.  Beyond this, to widths of around 150$^o$, typical SEP intensities rise by around three orders of magnitude.  There is then a further $\sim$order of magnitude increase out to ``full halo" CMEs, but the CME widths are likely to be unreliable.   The January 16, 2012 ``full halo" CME discussed above is circled.  

Above we noted that SEEDS CME speeds and widths were uncorrelated.  Nevertheless, the widths do show a similar correlation with SEP intensity as other catalogs (Figure~\ref{CMEwidthsep}(d)), with narrow CMEs, with widths less than  $\sim60^o$ typically associated with small SEP events, and large SEP events associated with large inferred CME widths.  DONKI and CORIMP CME widths in Figures~\ref{CMEwidthsep}(e) and (f) show weaker correlations with SEP intensity ($cc=0.438$ and 0.307, respectively).  As discussed above, CORIMP CME intervals may contain multiple structures so the CME width quoted in the catalog may not necessarily characterize the specific CME associated with the SEP event.

\begin{figure}    
 \centerline{\includegraphics[width=1.0\textwidth,clip=]{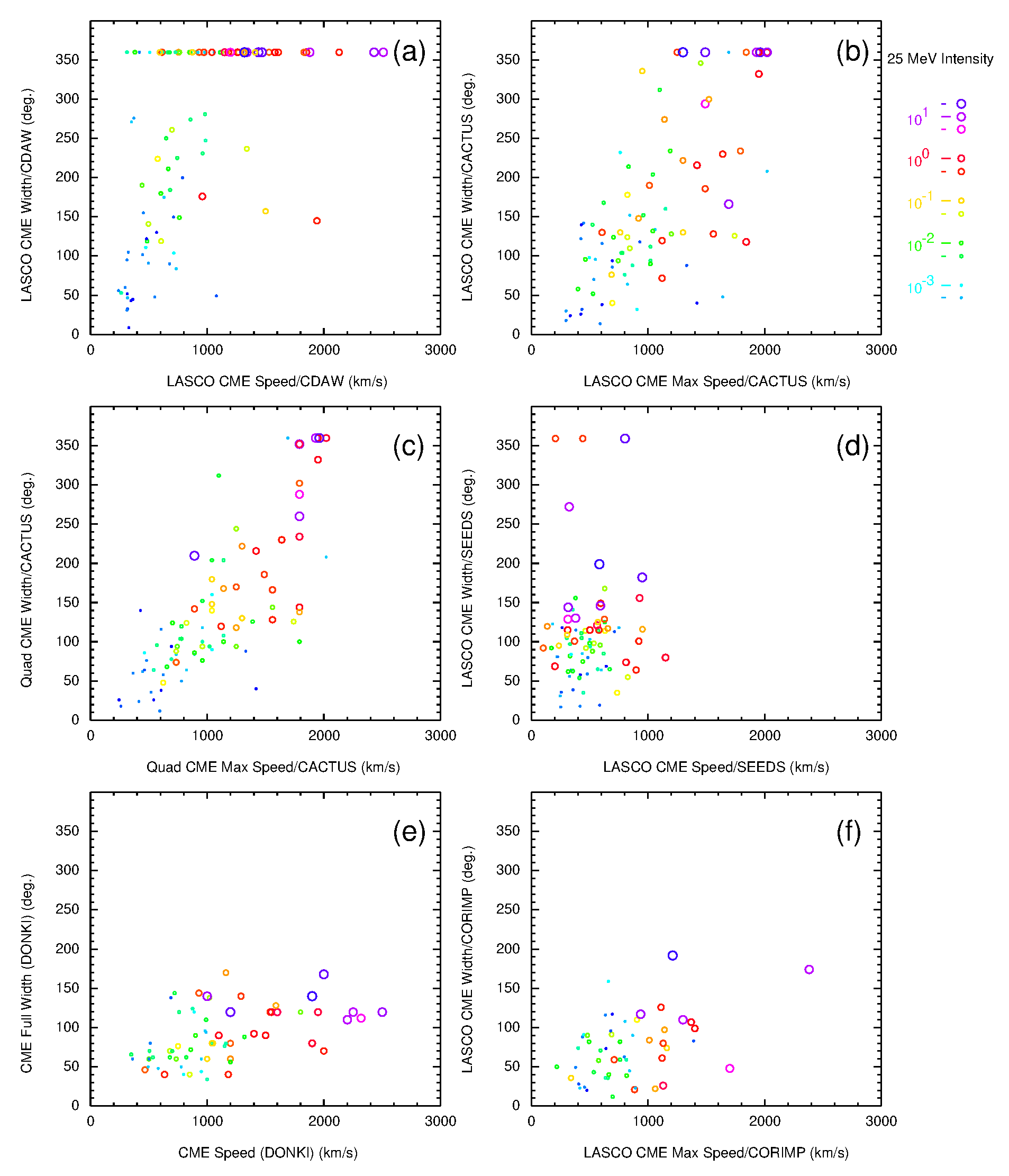}}
              \caption{Comparison of CME speeds \textit{vs.} widths in various catalogs with the 25 MeV proton intensity indicated by the color and size of the symbols, as indicated by the key, for events 30--90$^o$ west of the observing spacecraft.  }
   \label{CMEwvin}
   \end{figure}

Figure~\ref{CMEwvin} combines the results in Figures~\ref{widthspeed} and  \ref{CMEwidthsep} by showing the CME width plotted against CME speed for various catalogs, with the symbol size/color indicating the intensity of the associated 25~MeV proton events (see the scale on the right-hand side of the figure).  Again, events originating 30--90$^o$ west of the SEP-detecting spacecraft are included.  The overall conclusion from the figure is that the relationship between CME speed, width and SEP event intensity is highly dependent on the catalog used for the CME parameters.  However, there are some general features.  In particular, with the exception of SEEDS, the weakest SEP events tend to lie towards the bottom left corner of each plot, associated with slower, narrower CMEs.  Then there is a trend towards higher SEP intensities on moving towards the top and right-hand side of the plot, as CME speeds and widths increase.  This trend is particularly visible when quadrature spacecraft speeds and widths are used (c).  Since these parameters are expected to be less influenced by projection (and with the caveat again that the CACTUS parameters from different spacecraft may not be exactly equivalent), these results do suggest that more intense SEP events tend to associated with faster CMEs which also tend to also be wider, whereas weak SEP events are associated with slower, narrower CMEs.    

Figure~\ref{CMEwvin} summarizes the problems entailed with understanding how CME speed and width are related to SEP intensity using CME parameters in different catalogs.  The relationships are highly dependent on the catalog used.  Although there are general trends relating CME speed and width with SEP intensity, the values for individual events vary between catalogs.  There are also catalogs that do not report extremely wide CMEs, such as DONKI and CORIMP though intensity dependences on CME speed and width are still evident. 
      
\section{Summary and Discussion}

Although differences in the CME parameters in different catalogs have already been noted in a number of studies, the emphasis in this paper is on illustrating this for the CMEs associated with a recently reported set of SEP events that include 25 MeV protons. 
The catalogs considered were the CDAW, CACTUS, SEEDS, and CORIMP catalogs based on observations from the LASCO coronagaphs on SOHO, the CACTUS catalog produced using observations from STEREO-A and -B, and the DONKI catalog compiled by the CCMC.  Other inter-catalog comparisons have typically only used two catalogs, and to our knowledge, CORIMP and DONKI parameters have not been compared with a large sample of SEP event intensities. 

 We find that CME speeds in the different catalogs tend to be correlated as a group, but there can be large, often hundreds of km~s$^{-1}$, differences in the speeds of individual events. No single catalog appears to be more consistently reliable:  For some CMEs, CDAW parameters derived manually may be more representative than those in the CACTUS catalog, while for other CMEs, the opposite is the case.  CACTUS cannot estimate the speeds of the fastest CMEs that may be associated with the largest SEP events.  CME speeds reported by SEEDS rarely exceed 1000~km~s$^{-1}$.  CORIMP CME event intervals may include more than one CME, and hence the cataloged parameters may not characterize the specific CME that is associated with an SEP event.  We have also selected the more reliable speeds from either the CDAW or CACTUS LASCO catalogs, and used the CME speed from the spacecraft in quadrature to the solar event, reducing projection effects, but find that the CME speed -- SEP intensity correlation is little changed.   

Reported CME widths also are highly variable between catalogs.  The CDAW catalog indicates a high ($\sim50$\%) association between the 25 MeV proton events and ``full halo" ($360^o$-width) CMEs, but this association is not evident in other catalogs, where fewer such CMEs are identified.  DONKI, SEEDS and CORIMP  CME widths rarely exceed $\sim150^o$ though is some cases, SEEDS and CORIMP do not identify the full front of an extended CME.  Quadrature spacecraft observations suggest that the proton events are associated with a range of CME widths, with evidence of a correlation between SEP intensity and CME width, and also speed.  Proton event intensities increase rapidly, by $\sim3$ orders of magnitude, as CME widths increase above $\sim100^o$. 
 
Even if CME speeds and widths could be obtained reliably from single coronagraph observations, they are still in the plane of the sky and suffer from projection effects.  Improved estimates of CME parameters may be obtained by fitting observations from SOHO and STEREO to three-dimensional CME models (e.g., \opencite{ther11}, \opencite{theret11}, \opencite{liew11}, \opencite{bos12}, \opencite{r12},  \opencite{col13}) though at present, an extensive catalog of three-dimensional CME fits that may be compared with the properties of a large sample of SEP events does not yet exist. However, CME propagation speeds, even if more accurately determined, will remain crude proxies for the shock speeds at the field line connection point.  Thus, it is highly likely that, as for the catalogs used here, the overall correlation between CME speed and SEP intensity will be evident, but individual events will once again shuffle around in position as speed estimates are derived by researchers using various fitting techniques and interpreting coronagraph features differently.         
   
%
\begin{acks}
We thank the many individuals who have contributed to the development of the CME catalogs used in this study. The LASCO CME catalog is compiled at the CDAW Data Center by NASA and The Catholic University of America in cooperation with the Naval Research Laboratory.  The CACTUS CME catalog is maintained by the Solar Influences Data Analysis Center at the Royal Observatory of Belgium.   SEEDS is compiled at the Space Weather Laboratory of George Mason University and is supported by the NASA Living With a Star Program and NASA Applied Information Systems Research Program.  The Institute for Astronomy of the University of Hawaii produces the CORIMP catalog.  DONKI is developed at the Community Coordinated Modeling Center, NASA Goddard Space Flight Center.  SOHO is a project of international cooperation between ESA and NASA. We thank Leila Mays for information about the DONKI database.  This work was supported by the NASA Living With a Star Program as part of the activities of the Focussed Science Team studying the variability of solar energetic particle events.   
\end{acks}


\end{article} 
\end{document}